\documentclass[review]{elsarticle}

\usepackage{lineno,hyperref}
\usepackage{bm}
\usepackage{mathrsfs}
\usepackage{amssymb}
\usepackage{amsmath}
\usepackage{graphicx}
\usepackage{color}
\modulolinenumbers[5]
\usepackage{geometry}
\journal{}

\begin{document}

\begin{frontmatter}

\title{A class of diffusion algorithms with logarithmic cost over adaptive sparse Volterra network\tnoteref{mytitlenote}}
\tnotetext[mytitlenote]{* Corresponding author at: School of Electrical Engineering, Southwest Jiaotong University, Chengdu, Sichuan, 610031, PR China.\\
	E-mail\;addresses: {lulu@my.swjtu.edu.cn\;(L.\;Lu), hqzhao\_swjtu@126.com\;(H.\;Zhao)}.\\
}

\author{Lu Lu$^{a}$, Haiquan Zhao$^{a*}$}
\address{a)School of Electrical Engineering, Southwest Jiaotong University, Chengdu, China.}

\begin{abstract}
In this paper, we present a novel class of diffusion algorithms that can be used to estimate the coefficients of sparse Volterra network (SVN). The development of the algorithms is based on the logarithmic cost and $l_0$-norm constraint. Simulations for Gaussian and impulsive scenarios are conducted to demonstrate the superior performance of the proposed algorithms as compared with the existing algorithms.
\end{abstract}

\begin{keyword}
Distributed adaptation \sep Volterra filter \sep Sparse \sep Logarithmic cost.
\end{keyword}

\end{frontmatter}

\linenumbers

\section{Introduction}
The Volterra filter has been widely used as a nonlinear system modelling tool with considerable success \cite{contan2013modified,zhao2015adaptive,lu2016adaptive}. However, such a filter becomes very computationally expensive when a large number of coefficients are required. A second order Volterra (SOV) filter was developed to cope with the enormous amount of computations needed to obtain acceptable errors \cite{contan2013modified,zhao2015adaptive,lu2016adaptive,lu2016improved}.

Diffusion algorithms are the method for estimating parameters over adaptive networks, whose nodes can collect noisy observations related to a certain parameter of interest \cite{sayed2014adaptive}. Recently, some diffusion algorithms have been proposed \cite{sayed2014adaptive,abdolee2014estimation}. These algorithms aimed at enhancing the linear estimation performance have been presented in the literature, but few algorithms aimed at enhancing the nonlinear estimation capability of diffusion algorithms have been investigated. Particularly, in \cite{chouvardas2016diffusion}, an interesting trial was attempted to nonlinear adaptive learning by employing the kernel adaptive filter. Unfortunately, the structure of this method grows linearly with the number of processed patterns, which prohibits its practical applications.

Motivated by these considerations, in this paper, we proposed a new diffusion algorithm for adaptive sparse Volterra network (SVN). The parameters of SVN at every node are sparse, i.e., only a small portion of the coefficients (called active coefficients) have large magnitude while the rest of the coefficients (called inactive coefficients) are close or equal to zero. The development of the algorithm is based on an innovative approach: the algorithms are introduced based on the minimization of cost functions with logarithmic dependence on the adaptation error, instead of minimizing the pth power error. Moreover, these algorithms with $l_0$-norm constraint are proposed to achieve improved performance for SVN identification. 

\section{Problem formulation}

Consider the problem of estimating Volterra coefficients from a network with $N$ sensor nodes. In each iteration $i$, each sensor node $k \in \{ 1,2,...,N\}$ has access to the realization of some zero-mean random process $\{d_k(i),{\bm u}_{k.i}\}$, where $\{\cdot\}$ represents a set, $\bm{u}_{k,i}$ is an regression vector with length $M$, and $d_k(i)$ is the desired signal. Suppose the measurements arising from the model
\begin{equation}
d_k(i) = {\bm u}_{k,i}{\bm w}^o + v_k(i)
\label{001}
\end{equation}
where ${\bm w}^o$ is the coefficients of the Volterra series model, and $v_k(i)$ is the measurement noise. Assume that ${\bm u}_{k,i}$ and $v_k(i)$ are spatially independent and independent identically distributed (i.i.d.), and $v_k(i)$ is independent of ${\bm u}_{k,i}$. The output of the SOV network can be expressed as
\begin{equation}
\begin{aligned}
y_k(i) =& {\bm w}^T{\bm u}_{k,i} \\
=& {\sum\limits_{{m_1} = 0}^{P - 1} {\bm h}_{k,1}(m_1) {\bm x}_k(i - m_1)} \\ 
&+ \sum\limits_{{m_1} = 0}^{P - 1} {\sum\limits_{{m_2} = {m_1}}^{P - 1} {{{\bm h}_{k,2}}(m_1,m_2)} {\bm x}_k(i - m_2){\bm x}(i - m_1)}  
\end{aligned}
\label{002}
\end{equation}
where $(\cdot)^T$ denotes the transposition, ${\bm x}_k(i)$ is the input data at node $k$, $h_r$ is the $r$th order Volterra kernel at node $k$, and $P$ is the length of Volterra system, $P + P(P + 1)/2 = M$. The expanded input vector $\bm{u}_{k,i}$ and the expanded coefficients vector $\bm w$ of SOV system are expressed by
\begin{equation}
{\bm u}_{k,i} = {\left[{{\bm x}_k}(i),...,{{\bm x}_k}(i - P + 1),{\bm x}_k^2(i),{{\bm x}_k}(i){{\bm x}_k}(i - 1),...,{\bm x}_k^2(i - P + 1) \right]^T} 
\label{003}
\end{equation}
\begin{equation}
{\bm w} = {\left[{{{\bm h}_{k,1}}(0),...,{{\bm h}_{k,1}}(P-1),{{\bm h}_{k,2}}(0,0),{{\bm h}_{k,2}}(0,1),...,{{\bm h}_{k,2}}(P-1,...,P-1)} \right]^T} 
\label{004}
\end{equation}

\section{Proposed algorithm}
To obtain an improved performance, we adopted the local cost function that has logarithmic dependence on the error \cite{sayin2014novel}, i.e.,
\begin{equation}
J_k^{loc1}({\bm w}) \buildrel \Delta \over = \sum\limits_{l \in N_k} {a_{l,k}E\left[ {F(e_{l,i}) - \frac{1}{\delta}\ln \left\{\delta F(e_{l,i})\right\}} \right]}
\label{005}
\end{equation}
where $E(\cdot)$ denotes the expectation, $\delta>0$, $N_k$ is the set of nodes with which node $k$ shares information (including $k$ itself), and the weighting coefficients $\{{a_{l,k}}\}$ are real, non-negative. The function $F(e_{l,i})$ denotes a function of the error signal. Using the steepest descent algorithm, we hold the steepest descent adaptation as
\begin{equation}
{\bm w}_{k,i} = {\bm w}_{k,i-1} - \mu \sum\limits_{l \in N_k} {a_{l,k}\frac{\partial \left\{F(e_{l,i}) - \frac{1}{\delta}\ln \left({\delta F(e_{l,i})} \right) \right\}}{\partial \bm w}|_{\bm w_{k,i-1}} } 
\label{006}
\end{equation}
where $\mu$ is the step size. Under the linear combination assumption \cite{sayed2014adaptive}, let us define the linear combination $\bm w_{k,i}$ at node $k$ as  ${\bm w}_{k,i-1} = \sum\limits_{l \in N_k} {{a_{l,k}}{\bm \varphi_{l,i-1}}}$, where ${\bm \varphi}_{k,i}$ is the local estimates at node $k$. Moreover, in adaptation step, ${\bm \varphi}_{k,i-1}$ is replaced by linear combination ${\bm w}_{k,i-1}$. Such substitution is reasonable, since the linear combination contains more data information from neighbour nodes than ${\bm \varphi}_{k,i-1}$ \cite{abdolee2014estimation}. Then, we can obtain the iterative expressions for the proposed algorithms.

The diffusion least mean logarithmic square (dLMLS) algorithm

For $F(e_{l,i}) = E(e_{l,i}^2)$ this simplifies into the dLMLS algorithm:
\begin{equation}
{\bm \varphi}_{k,i} = {\bm w}_{k,i-1} + \mu \sum\limits_{l \in N_k} c_{l,k} \frac{\delta {\bm u}_{l,i}^Te_{l,i}^3}{1 + \delta e_{l,i}^2}
\label{007}
\end{equation}
where $\{{c_{l,k}}\}$ is the non-negative weighting coefficients, satisfying the condition $c_{l,k} = a_{l,k} = 0$ if $l \notin {N_k}$.

The diffusion least logarithmic absolute difference (dLLAD) algorithm

The dLLAD algorithm is derived by substituting $F(e_{l,i}) = E(|e_{l,i}|)$ in the general formula given by (6). Its update is given by
\begin{equation}
{\bm \varphi}_{k,i} = {\bm w}_{k,i-1} + \mu \sum\limits_{l \in N_k} c_{l,k} \frac{\delta {\bm u}_{l,i}^T{e_{l,i}}}{1 + \delta |e_{l,i}|}
\label{008}
\end{equation}
where $|\cdot|$ denotes absolute value of a scalar.

The diffusion logarithmic least mean $p$-power (dLLMP) algorithm

For $F(e_{l,i}) = E(|e_{l,i}|^{p_{l,i}})$, we obtain
\begin{equation}
{\bm \varphi}_{k,i} = {\bm w}_{k,i-1} + \mu \sum\limits_{l \in N_k} c_{l,k} \frac{\delta {\bm u}_{l,i}^T|e_{l,i}|^{2p_{l,i} - 1}sign(e_{l,i})}{1 + \delta |e_{l,i}|^{p_{l,i}}}
\label{009}
\end{equation}

Remark 3.1: The proposed dLMLS algorithm is based on fourth-statistics of the error, and can therefore achieve a smaller steady-state kernel error. The dLLAD algorithm intrinsically combines the $l_2$-norm and $l_1$-norm. In impulsive noise environments, it may be expected to converge faster than the dLMS algorithm does.

Remark 3.2: The dLLMP algorithm resembles the algorithm that we proposed in \cite{lu2016improved}. Because the logarithmic-order class includes $\alpha$-stable noise process, it can be used estimate the coefficient with reduced negative effects of outliers. When $p_{l,i}=1$, it reduces to the dLLAD algorithm. When $p_{l,i}=2$, it becomes the dLMLS algorithm.

Remark 3.3: For the Eq. (7-9) above, the diffusion algorithms obtain solutions via Adapt-then-Combine (ATC) step. The Combine-then-Adapt (CTA) \cite{sayed2014adaptive} algorithms can be easily derived by exchanging the order of these two steps.

$l_0$-norm-based algorithms: In diffusion adaptation, one often encounters many systems with sparsity property, i.e., there are only a small number of nonzero entries in the impulse response at every node. To introduce sparsity, we can minimize the following penalized local cost function with $l_0$-norm:
\begin{equation}
J_k^{loc2}(\bm w) \buildrel \Delta \over = \sum\limits_{l \in N_k} {a_{l,k}E\left[ {F(e_{l,i}) - \frac{1}{\delta}\ln \left({\delta F(e_{l,i})} \right)} \right] + \lambda ||\bm w||_0}
\label{010}
\end{equation}
where $||\cdot||_0$ denotes the $l_0$-norm, $\lambda$ is a controller factor to balance the new penalty and the estimation error. Similarly, the adaptation of logarithmic cost with $l_0$-norm can be given as
\begin{equation}
\left\{\begin{array}{l}
{\bm \varphi}_{k,i} = {\bm w}_{k,i-1} - \mu \sum\limits_{l \in N_k} {c_{l,k}\nabla F({e_{k,i}})\frac{\delta F(e_{k,i})}{1 + \delta F(e_{k,i})} - \rho \nabla ||{\bm w}_{k,i-1}||_0}  \\ 
{\bm w}_{k,i} = \sum\limits_{l \in N_k} {a_{l,k}{\bm \varphi}_{l,i}}  \\ 
\end{array} \right.
\label{011}
\end{equation}
where $\rho=\lambda\mu$. Since the minimization of the $l_0$-norm is a Non-Polynomial (NP) hard problem, $\nabla ||{\bm w}_{k,i-1}||_0$ is usually approximated by a first-order Taylor series expansion, i.e.,
\begin{equation}
\nabla ||{\bm w}_m||_0 = \left\{\begin{array}{l}
- \beta^2{\bm w}_m - \beta,\;\;\;\;\;\;-1/\beta \le {\bm w}_m < 0 \\ 
- \beta^2{\bm w}_m + \beta,\;\;\;\;\;\;0 < {\bm w}_m \le 1/\beta \\ 
0,\;\;\;\;\;\;\;\;\;\;\;\;\;\;\;\;\;\;\;\;\;\;\;\;elsewhere \\ 
\end{array} \right.
\label{012}
\end{equation}
where $\beta \in [5,20]$ is the positive constant. It should be noted that only $\bm w$ within the neighborhood of zero, i.e., $[- 1/\beta ,1/\beta]$ which is named \emph{attraction region}, are attracted. Besides, in this region, the closer $\bm w$ is to zero, the greater the attraction intensity is. If $\bm w$ is not in this region, no attraction will be performed. We can conclude that based on the \emph{zero-attraction} function in (12), the $l_0$-based algorithms can enhance the performance of diffusion algorithm for SVN, because in such network the near-zero coefficients are dominant. Moreover, a larger $\beta$ results in a stronger intensity but a narrower attraction region, which affects the performances of the algorithms.

\section{Simulation results}

We present the simulation results to verify the effectiveness of the proposed algorithms in comparison with the diffusion LMS (dLMS) \cite{sayed2014adaptive}, and the diffusion LMP (dLMP) algorithm \cite{wen2013diffusion}. We consider a SVN composed of 20 nodes. The unknown plant which is given in \cite{wen2013diffusion} is modelled as a second-order nonlinear system with $M$=14 ($P$=4). The Gaussian signal with zero mean and unit variance is employed as the exciting input. In the simulation study, the effectiveness is assessed in terms of network mean-square deviation (NMSD), which is defined as 
\begin{equation}
\mathrm{NMSD} \buildrel \Delta \over = \frac{1}{N}\sum\limits_{k=1}^N {||\bm w_k - \bm w^o||_2^2}
\label{013}
\end{equation}
The noise is modelled by standard symmetric $\alpha$-stable ($S\alpha S$) distribution $\eta(t)\buildrel \Delta \over =\mathrm{exp}\left(-|t|^\alpha\right)$ \cite{lu2016improved}, where $\alpha$ is a characteristic exponent. In all our simulations, the curves are drawn from the average of 25 independent runs.

In first example, the noise signal is generated with $\alpha=2$, which corresponds to the Gaussian distribution. Fig. 1 illustrates the NMSD curves for algorithms. It can be observed that the proposed algorithms outperform the dLMS algorithm. The only sacrifice to make is a slow convergence rate at the initial stage.

In second example, $\alpha$ of all nodes are ranging from 1.2 to 1.8. The learning curves for the algorithms are plotted in Fig. 2, where $p$ in dLLMP is chosen by random. We can clearly see that the dLLMP $l_0$ method is very robust to the impulsive noise and achieves small kernel estimation error. The other algorithms, in contrast, have large fluctuations during the adaptation. 

\begin{figure}[!htb]
	\centering
	\includegraphics[scale=0.5] {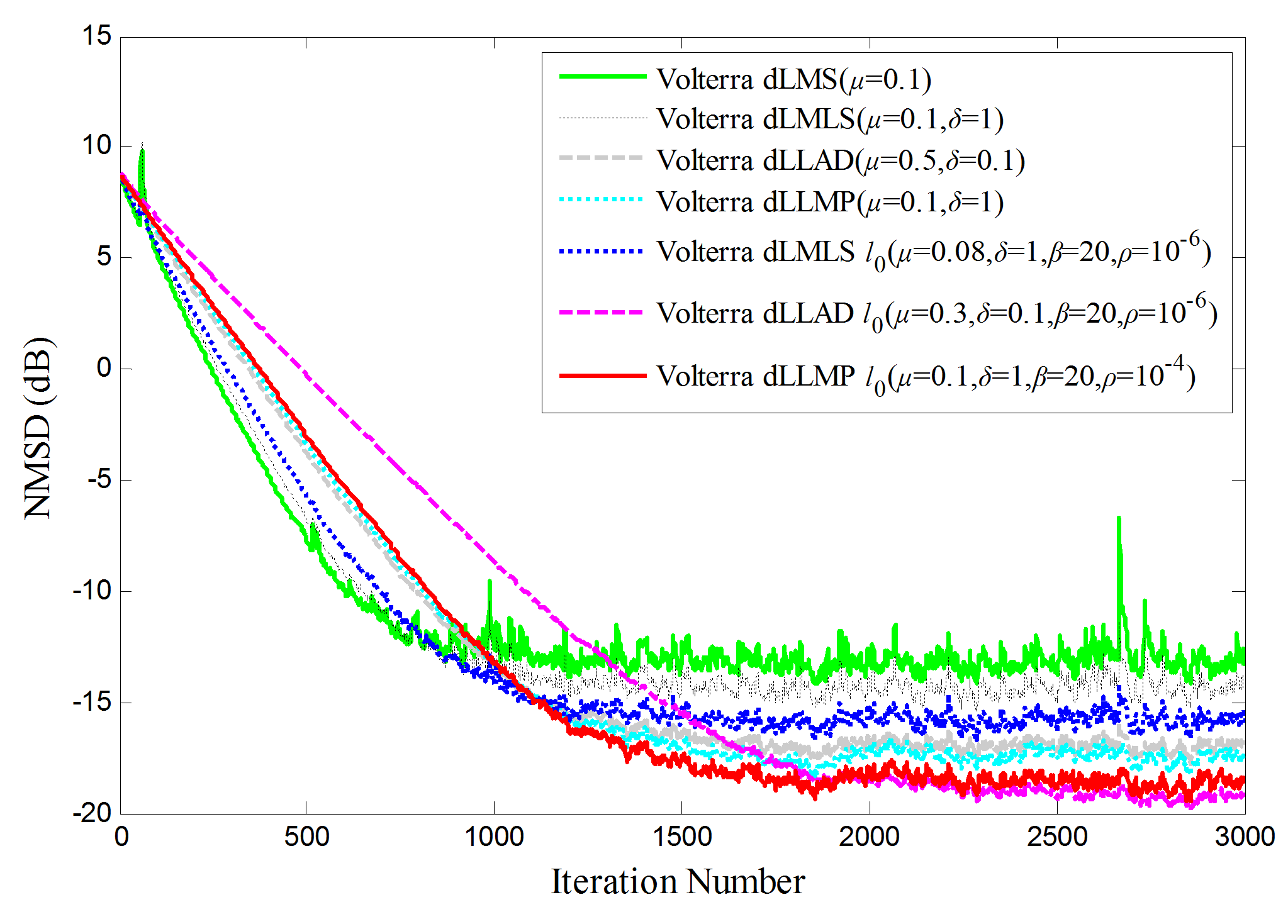}
	\caption{\label{1} NMSD in SVN for $\alpha=2$.}
	\label{Fig01}
\end{figure}

\begin{figure}[!htb]
	\centering
	\includegraphics[scale=0.5] {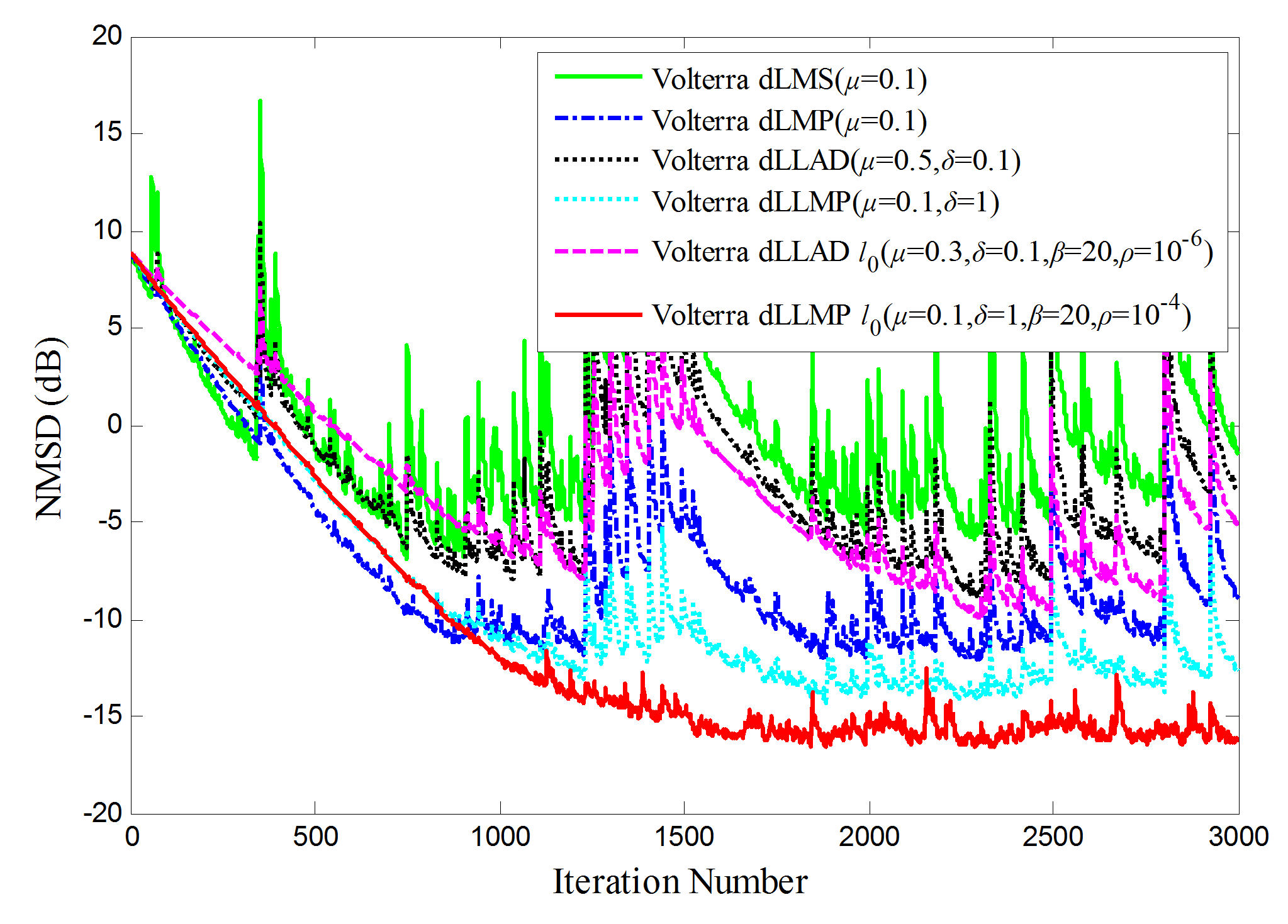}
	\caption{\label{1} NMSD in SVN for $S\alpha S$ noise.}
	\label{Fig02}
\end{figure}

\section{Conclusion}

An adaptive SVN using a class of logarithmic cost algorithms has been proposed and investigated in detail through simulations. In Gaussian noise environment, the proposed algorithms enjoy smaller kernel misadjustment as compared with the dLMS algorithm. If the network is corrupted by α-stable noise, the proposed dLLMP algorithm will provide overwhelmingly better stability in comparison with the existing algorithms.

\section*{Acknowledgments}
This work was partially supported by National Science Foundation of P.R. China (Grant: 61571374, 61271340, 61433011). The first author would also like to acknowledge the China Scholarship Council (CSC) for providing him with financial support to study abroad (No. 201607000050).

\bibliography{mybibfile}

\end{document}